\documentclass[
  aps,
  prl,
  reprint,
  amsmath,
  amssymb,
  superscriptaddress,
  nofootinbib,
  floatfix
]{revtex4-2}

\usepackage{graphicx}
\usepackage{bm}
\usepackage{booktabs}
\usepackage{microtype}
\usepackage{hyperref}
\usepackage{xcolor}
\usepackage{comment}
\usepackage{multirow} 
\usepackage{amsmath}

\hypersetup{
  colorlinks=true,
  linkcolor=blue,
  citecolor=blue,
  urlcolor=blue
}

\begin{document}

\title{Tight bound on nanohertz Gravitational-Wave Background with 17.6 years of Fermi-LAT data}

\author{Zi-Qing Xia}
\email{xiazq@pmo.ac.cn}
\affiliation{Key Laboratory of Dark Matter and Space Astronomy, Purple Mountain Observatory, Chinese Academy of Sciences, Nanjing 210033, China}

\author{Xiaoyuan Huang}
\affiliation{Key Laboratory of Dark Matter and Space Astronomy, Purple Mountain Observatory, Chinese Academy of Sciences, Nanjing 210033, China}
\affiliation{School of Astronomy and Space Science, University of Science and Technology of China, Hefei, Anhui 230026, China }

\author{Qiang Yuan}
\email{yuanq@pmo.ac.cn}
\affiliation{Key Laboratory of Dark Matter and Space Astronomy, Purple Mountain Observatory, Chinese Academy of Sciences, Nanjing 210033, China}
\affiliation{School of Astronomy and Space Science, University of Science and Technology of China, Hefei, Anhui 230026, China }

\author{Yi-Zhong Fan}
\email{yzfan@pmo.ac.cn}
\affiliation{Key Laboratory of Dark Matter and Space Astronomy, Purple Mountain Observatory, Chinese Academy of Sciences, Nanjing 210033, China}
\affiliation{School of Astronomy and Space Science, University of Science and Technology of China, Hefei, Anhui 230026, People's Republic of China}

\date{\today}

\begin{abstract}
With 17.6 yr of Fermi-LAT Gamma-ray Pulsar Timing Array (GPTA) data, spanning October 2008 to June 2026, we construct four datasets with different observing cadences and perform Bayesian analyses to search for a stochastic gravitational-wave background (GWB) with the Hellings-Downs spatial correlation pattern, adopting a fixed spectral index of $\gamma=13/3$, corresponding to the standard supermassive black hole binary background. All four datasets yield Bayes factors close to unity, indicating no strong statistical preference for either the GWB model or the null hypothesis. We therefore  place a 95\% upper limit on the strain amplitude of $A_{\rm GWB}<3.72\times10^{-15}$ at a reference frequency of $1\,{\rm yr}^{-1}$, using the 122-day observing cadence. This improves the previous GPTA limit from the 12.5 yr dataset by a factor of approximately 2.7, 
and challenges the GWB amplitude reported by MeerKAT PTA in 2025. 
The projected time for the GPTA to reach the common amplitude range inferred by NANOGrav/EPTA/PPTA/CPTA experiments in 2023 is brought forward from 2030 to mid-2027. Our results demonstrate the rapidly growing sensitivity of GPTA and its key role as a complementary probe of the nanohertz GWB.
\end{abstract}

\maketitle

\textit{Introduction.--}
Pulsar timing arrays (PTAs) probe gravitational waves (GWs) in the nanohertz frequency band by monitoring the pulse arrival times of an ensemble of highly stable millisecond pulsars (MSPs). A stochastic gravitational-wave background (GWB) induces correlated fluctuations in pulsar timing residuals. The resulting angular correlation pattern is described by the Hellings--Downs (HD) curve, as expected for an isotropic background of tensor GWs \cite{HellingsDowns1983,FosterBacker1990}. The detection and characterization of an HD-correlated GWB offer a unique opportunity to study the population of supermassive black hole (SMBH) binaries and the evolution of massive black holes across cosmic time \cite{Sesana2013SMBHB,BurkeSpolaor2019PTAReview,Shannon2013PPTA}.

The scientific reach of PTAs extends well beyond the search for a GWB. Errors in the Solar-system ephemeris (SSE) affect the transformation of observatory arrival times to the Solar-system barycenter, and inaccuracies in planetary masses or orbits can therefore produce timing perturbations correlated across pulsars. The resulting dipolar correlation pattern enables PTA observations to constrain the masses and dynamical properties of Solar-system bodies and to test alternative ephemerides \cite{Champion2010PlanetMasses,Guo2019SolarSystemDynamics,Caballero2018SolarSystem,2026arXiv260929162W}. PTAs also provide a means of probing dark matter. For example, ultralight scalar, vector, or tensor fields can generate time-dependent gravitational potentials or induce couplings to ordinary matter, leaving characteristic signatures in pulsar timing residuals \cite{KhmelnitskyRubakov2014,Porayko2018ULDM,Xue2022DarkPhoton,Wu2022VectorDM,FermiLATULDM2023,Luu2024GPTAWaveDM,KimMitridate2024,Unal2024,BoddyDrorLam2025}. These applications illustrate the broader role of pulsar timing as a precision tool for testing gravitational physics and probing dark matter candidates.

Several radio PTA collaborations, including the European PTA (EPTA) \cite{vanHaasteren2011EPTA,Lentati2015EPTA,Chen2021EPTA,EPTAInPTA2023}, the Parkes PTA (PPTA) \cite{Jenet2006PPTA,Shannon2013PPTA,Shannon2015PPTA,Goncharov2021PPTA,Reardon2023PPTA}, NANOGrav \cite{Demorest2013NANOGrav,Arzoumanian2016NANOGrav,Arzoumanian2018NANOGrav,NANOGrav2020,NANOGrav2023}, the International PTA (IPTA) \cite{Verbiest2016IPTA,Antoniadis2022IPTA}, the Chinese PTA (CPTA) \cite{CPTA2023}, and the MeerKAT PTA (MPTA) \cite{Miles2025MPTA}, have carried out long-term pulsar timing observations, steadily advancing the search for nanohertz gravitational waves. Recent data releases have reported evidence for a common-spectrum stochastic process, with spatial correlations broadly consistent with the HD signature of a GWB \cite{NANOGrav2020,Chen2021EPTA,Goncharov2021PPTA,Antoniadis2022IPTA,EPTAInPTA2023,Reardon2023PPTA,NANOGrav2023,CPTA2023,Miles2025MPTA}. Complementing these radio-based efforts, recent NICER observations have enabled independent searches for the nanohertz GWB in X-rays, placing new constraints despite the absence of statistically significant evidence for the background \cite{Cao2026NICER}. These complementary efforts across different observational wavebands provide independent avenues for probing the nanohertz gravitational-wave background.

The Fermi Large Area Telescope (Fermi-LAT), a pair-conversion telescope aboard the Fermi Gamma-ray Space Telescope, provides an alternative approach to pulsar timing through the detection of gamma-ray photons \cite{Atwood2009LAT,Ackermann2012LATPerformance}. Unlike radio observations, gamma-ray timing is unaffected by interstellar dispersion, providing an independent timing data set with distinct observational systematics. The third Fermi-LAT catalog of gamma-ray pulsars has substantially expanded the sample available for long-term timing studies \cite{Fermi3PC2023}. Photon-probability weighting enhances the sensitivity of pulsation analyses by accounting for the probability that each detected photon originated from the target pulsar \cite{Kerr2011PhotonWeights}. In addition, maximum-likelihood and template-based techniques have been developed to derive precise gamma-ray times of arrival (TOAs) from photon data \cite{Ray2011GammaTiming}.

Gamma-ray MSPs have also been used to search for the nanohertz GWB. In 2022, the Fermi-LAT Collaboration established a gamma-ray pulsar timing array (GPTA) using 12.5 years of observations of 35 bright gamma-ray pulsars, placing a 95\% credible upper limit of $1.0\times10^{-14}$ on the characteristic strain at a reference frequency of $1\,\mathrm{yr}^{-1}$ \cite{FermiGPTA2022}. Subsequent work has used the GPTA to constrain ultralight dark-matter scenarios \cite{FermiLATULDM2023,Luu2024GPTAWaveDM}, while subsequent studies have further explored the potential of GPTA for GWB searches and improved the sensitivity of GPTA\cite{Xie2023GPTA,Valtolina2026}. Future space-based gamma-ray observatories could detect thousands of MSPs, potentially enhancing the GPTA sensitivity to the nanohertz GWB to levels comparable to or exceeding those of current-generation radio PTAs \cite{Kerr2026FutureGPTA}. Together, these efforts highlight the GPTA as an independent yet complementary counterpart to radio PTAs, providing opportunities for multiwavelength studies of nanohertz GWs and dark-matter candidates.

In this work, we analyze approximately 17.6 years of Fermi-LAT observations and construct four GPTA data sets with different observing cadences. We perform Bayesian searches for a stochastic GWB and derive constraints on its characteristic strain amplitude. By comparing the GWB constraints obtained from these data sets, we evaluate the sensitivity of the GPTA at different observing cadences and discuss possible improvements to future gamma-ray pulsar timing observations.

\textit{Fermi-LAT Data and Analysis.--}
We analyze publicly available Fermi-LAT photon data \cite{FSSCDataAccess} collected between 2008 October 27 and 2026 June 13, spanning approximately 17.6 years. 
Our initial sample consists of the 35 bright gamma-ray MSPs used in the first GPTA analysis \cite{FermiGPTA2022}. 
For each pulsar, we select photons in the energy range of 100\,MeV--100\,GeV within $3^\circ$ of its position. The analysis uses the Pass 8 event reconstruction and the corresponding LAT instrument response functions (IRFs) \cite{Ackermann2012LATPerformance,Atwood2013Pass8,FSSCPass8}. 
We select SOURCE-class events with the FRONT+BACK conversion type. To reduce contamination from gamma-ray emission originating from the Earth's limb, we exclude events with zenith angles greater than $100^\circ$. We also apply the recommended quality cuts, \texttt{DATA\_QUAL>0 \&\& LAT\_CONFIG==1}, to define the good time intervals.
Photon probabilities are calculated using the source and background models from the Fourth Fermi-LAT Source Catalog (4FGL)\cite{Abdollahi2020Fermi4FGL} and are used to weight individual events according to their probabilities of originating from the target MSP \cite{Kerr2011PhotonWeights}.
The photon data selection and preparation described above are performed using the Fermi Science Tools \cite{FermiScienceTools}.

For each pulsar, photon phases are calculated using the corresponding timing model adopted from the first GPTA analysis \cite{FermiGPTA2022}. Following the methodology described in Ref.~\cite{FermiGPTA2022}, we then construct a pulse-profile template and use it to estimate gamma-ray times of arrival (TOAs) from the weighted photon data. The resulting TOAs are then used to obtain timing residuals through the timing-model refitting. Adopting the DE440 Solar-system ephemeris \cite{Park2021DE440}, we use PINT for the timing-model construction and refitting \cite{Luo2021PINT}.

To investigate the sensitivity of the GPTA at different observing cadences, we construct four data sets with cadences of 122, 182, 243, and 365 days. For each data set, we divide the observation span into intervals according to the specified cadence and derive one TOA from each interval. The final interval is also included, even if its duration is shorter than the nominal cadence. These data sets contain 21, 25, 29, and 32 MSPs, respectively, with 53, 36, 27, and 18 TOAs per pulsar. The properties of these data sets and the corresponding GWB constraints are summarized in Table~\ref{tab:gwb_results}. Details of TOA estimation are provided in the Supplemental Material.

\textit{GWB Search and Results.--}
We search the GPTA data for a common-spectrum stochastic red process, which could be associated with the GWB. The timing-residual covariance model accounts for both individual-pulsar noise and common stochastic signals. The noise of each pulsar is modeled with white-noise and intrinsic red-noise components. The white-noise component accounts for uncorrelated TOA measurement uncertainties and any additional scatter beyond the nominal measurement errors. In gamma-ray pulsar timing, the white-noise level is largely governed by the finite number of detected photons, which determines the precision of the TOA measurements. The intrinsic red noise in gamma-ray pulsar timing is primarily associated with spin noise, which accounts for long-timescale stochastic variations in the pulsar's rotational phase and is generally modeled with a power-law PSD (i.e., power spectral density). Unlike radio timing, gamma-ray timing is unaffected by interstellar dispersion and therefore does not suffer from chromatic timing noise induced by variations in the dispersion measure. The noise analysis and subsequent GWB search are performed using \texttt{enterprise} \cite{Enterprise2020}, with \texttt{PTMCMCSampler} \cite{Ellis2017PTMCMCSampler} employed to sample the posterior distributions of the model parameters. Details of individual-pulsar noise analyses are provided in the Supplemental Material.

The stochastic GWB produced by a population of SMBH binaries is described by a power-law characteristic strain spectrum,
\begin{equation}
h_{\rm c}(f)=A_{\rm GWB}\left(\frac{f}{f_{\rm yr}}\right)^{\alpha},
\label{eq:hc}
\end{equation}
where $A_{\rm GWB}$ is the characteristic strain amplitude at the reference frequency $f_{\rm yr}=1\,\mathrm{yr}^{-1}$, and $\alpha$ is the spectral index. For the standard SMBH binary GWB, $\alpha=-2/3$.
Following the standard PTA formalism \cite{Phinney2001StochasticBG,RomanoCornish2017}, the corresponding PSD of the induced timing residuals is given by
\begin{equation}
P_{\rm GWB}(f)=\frac{A_{\rm GWB}^{2}}{12\pi^{2}}\left(\frac{f}{f_{\rm yr}}\right)^{-\gamma}f_{\rm yr}^{-3},
\label{eq:power}
\end{equation}
where the spectral indices of the characteristic strain and timing-residual PSD are related by $\gamma=3-2\alpha=13/3$ for a population of SMBH binaries. The spatial correlations induced by an isotropic GWB are described by the HD function.

\begin{table}[b]
\caption{Properties and GWB constraints of the four GPTA data sets. All analyses include BayesEphem, with the timing-residual spectral index fixed at $\gamma=13/3$. $N_{\rm TOA}$ denotes the number of TOAs per pulsar. The quoted $\log_{10} A_{\rm GWB}^{95\%}$ values represent 95\% posterior upper limits on $\log_{10} A_{\rm GWB}$.}
\label{tab:gwb_results}
\begin{ruledtabular}
\begin{tabular}{ccccc}
Cadence (days) &
$N_{\rm PSR}$ &
$N_{\rm TOA}$ per MSP &
$\log_{10} A_{\rm GWB}^{95\%}$ &
$\mathcal{B}$ \\
\hline
122 & 21 & 53 & $-14.43$ & 0.65 \\
182 & 25 & 36 & $-14.30$ & 0.69 \\
243 & 29 & 27 & $-14.24$ & 0.73 \\
365 & 32 & 18 & $-14.36$ & 0.66 \\
\end{tabular}
\end{ruledtabular}
\end{table}

\begin{figure*}
    \centering
    \includegraphics[width=0.98\textwidth]{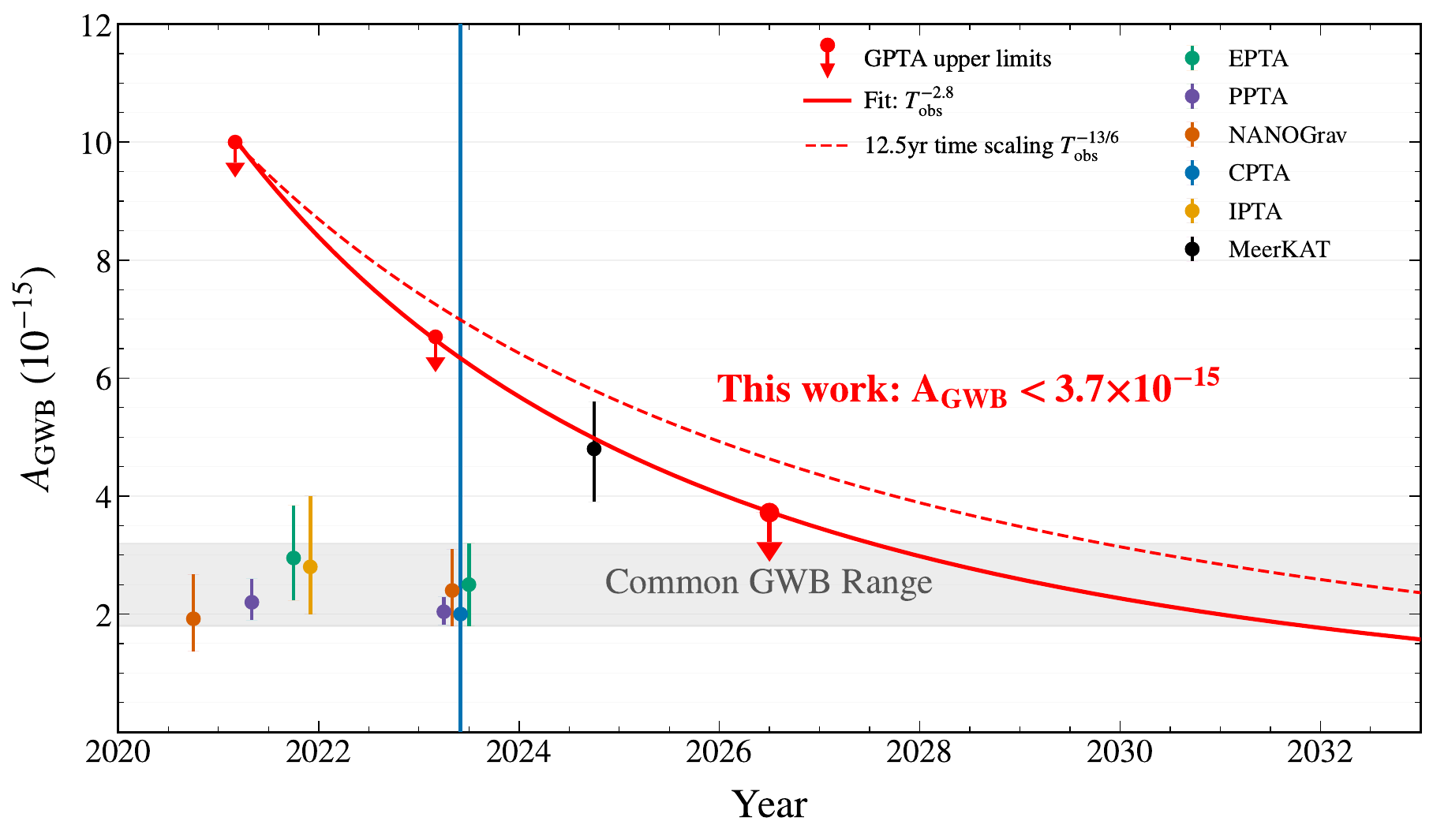}
    \caption{Comparison of the GWB amplitude measurements reported by different radio PTAs and the upper limits obtained by the GPTA. The vertical axis shows the characteristic strain amplitude \(A_{\rm GWB}\), in units of \(10^{-15}\), assuming a fixed spectral index of \(\gamma=13/3\), at a reference frequency of \(1\,{\rm yr}^{-1}\). Circular markers with error bars represent amplitude measurements and their associated uncertainties from the EPTA (green; \cite{Chen2021EPTA,EPTAInPTA2023}), PPTA (purple; \cite{Goncharov2021PPTA,Reardon2023PPTA}), NANOGrav (brown; \cite{NANOGrav2020,NANOGrav2023}), IPTA (yellow; \cite{Antoniadis2022IPTA}), CPTA (blue; \cite{CPTA2023}), and MeerKAT PTA (MPTA; black; \cite{Miles2025MPTA}). The gray shaded region indicates the approximate common range suggested by the candidate GWB signals reported by radio PTAs in 2023 \cite{EPTAInPTA2023,CPTA2023,Reardon2023PPTA,NANOGrav2023}. The GPTA upper limits are highlighted in red: the 12.5-year DR1 limit (\(A_{\rm GWB}<10\times10^{-15}\); \cite{FermiGPTA2022}), the preliminary 14.5-year limit (\(A_{\rm GWB}<6.7\times10^{-15}\); \cite{Kerr2023GPTA}), and the 17.6-year upper limit obtained in this work (\(A_{\rm GWB}<3.72\times10^{-15}\)). The red solid curve shows a power-law fit to the three GPTA upper limits, \(A_{\rm GWB}\propto T_{\rm obs}^{-2.8}\), while the red dashed curve illustrates the expected sensitivity improvement under the weak-signal scaling \(A_{\rm GWB}\propto T_{\rm obs}^{-13/6}\) \cite{Siemens2013Scaling}, normalized to the GPTA DR1 result at 12.5 years \cite{FermiGPTA2022}.}
    \label{fig:GWB_PTA_comparison}
\end{figure*}

We analyze the four cadence data sets independently, deriving posterior distributions for the GWB amplitude and calculating the Savage--Dickey (SD) Bayes factors ($\mathcal{B}$) \cite{Dickey1971,VerdinelliWasserman1995,Trotta2007SavageDickey}. In the GWB search, we jointly fit the GWB model together with the white-noise and red-noise models for all MSPs, allowing all noise parameters to vary freely. We also include the BAYESEPHEM \cite{Vallisneri2020BayesEphem} model to account for potential systematic effects arising from Solar-system ephemeris uncertainties, using DE440 in our analysis \cite{Park2021DE440}.
The Bayes factors and 95\% upper limits are summarized in Table~\ref{tab:gwb_results}. The resulting Bayes factors are $\mathcal{B}=0.65$, $0.69$, $0.73$, and $0.66$ for the 122-, 182-, 243-, and 365-day data sets, respectively. All four values are below unity, indicating that the data do not favor the GWB signal model over the noise-only model. We therefore derive upper limits on the GWB amplitude rather than claim a detection.
The 122-day data set yields the most stringent 95\% upper limit, $A_{\rm GWB}<3.72\times10^{-15}$, as shown in Figure~\ref{fig:GWB_PTA_comparison}. 
The improved constraint relative to the sensitivity projection based on the 12.5-year GPTA DR1 result \cite{FermiGPTA2022} (red dashed curve in Figure~\ref{fig:GWB_PTA_comparison})  is primarily attributable to the shorter observing cadence, which enables more frequent sampling and yields a larger number of TOAs.
Although this strategy reduces the number of MSPs included in the analysis, particularly those with relatively low photon statistics, it allows more frequent TOA generation for brighter MSPs with higher timing precision, thereby increasing the number of high-quality TOAs.

To investigate alternative common processes, we consider common-spectrum red-noise models with different spatial correlation patterns, including uncorrelated, monopolar, and dipolar correlations, as well as HD correlations with a free timing-residual spectral index $\gamma$. 
We perform signal searches both with and without BAYESEPHEM. 
Details of these searches and the corresponding results are provided in the Supplemental Material.

\textit{Discussion.--} Using 17.6 years of Fermi-LAT observations, we searched for a nanohertz stochastic GWB with the GPTA. We constructed four data sets with different observing cadences and performed Bayesian analyses, adopting a power-law GWB spectrum with a fixed spectral index of $\gamma=13/3$, characteristic of a population of inspiralling SMBH binaries, and the HD spatial correlation pattern. For all four cadences, the Bayes factors were below unity ($\mathcal{B}=0.65$--$0.73$), indicating that the data do not favor the GWB signal model over the noise-only model. We therefore found no evidence for a GWB and derived a 95\% upper limit of $A_{\rm GWB}<3.72\times10^{-15}$, obtained with the 122-day cadence. This result provides an improved constraint on the nanohertz GWB using gamma-ray pulsar timing.

Figure~\ref{fig:GWB_PTA_comparison} compares our result with previous GPTA upper limits and GWB amplitude measurements reported by various radio PTAs. In 2023, several major radio PTA collaborations reported evidence for a common-spectrum process, with characteristic strain amplitudes of a few $10^{-15}$, as indicated by the gray region in Figure~\ref{fig:GWB_PTA_comparison}. Assuming a fixed spectral index of $\gamma=13/3$, the PPTA DR3 analysis reported an amplitude of $A_{\rm GWB}=2.04^{+0.25}_{-0.22}\times10^{-15}$ \cite{Reardon2023PPTA}, while the EPTA DR2 analysis obtained $A_{\rm GWB}=(2.5\pm0.7)\times10^{-15}$ \cite{EPTAInPTA2023}. Similarly, the NANOGrav 15-year data set yielded $A_{\rm GWB}=2.4^{+0.7}_{-0.6}\times10^{-15}$ \cite{NANOGrav2023}. The CPTA also reported a characteristic amplitude of approximately $2.0\times10^{-15}$ \cite{CPTA2023}. More recently, the MeerKAT PTA (MPTA) reported an amplitude of $A_{\rm GWB}=4.8^{+0.8}_{-0.9}\times10^{-15}$ for $\gamma=13/3$ \cite{Miles2025MPTA}.

The GPTA provides an independent probe of the nanohertz GWB through gamma-ray observations of millisecond pulsars. Unlike radio timing, gamma-ray timing is unaffected by dispersive propagation delays associated with the ionized interstellar medium, offering a complementary approach to radio PTAs. The GPTA DR1 analysis, based on 12.5 years of Fermi-LAT observation, placed a 95\% upper limit of $A_{\rm GWB}<10\times10^{-15}$ \cite{FermiGPTA2022}. This limit was subsequently improved to $A_{\rm GWB}<6.7\times10^{-15}$ using the preliminary 14.5-year DR2 data set \cite{Kerr2023GPTA}. With the 17.6-year data set analyzed here, we further improve the upper limit to $A_{\rm GWB}<3.72\times10^{-15}$, representing improvements by factors of approximately 2.7 and 1.8 relative to DR1 and preliminary DR2, respectively.
Notably, our 95\% upper limit is below the GWB amplitude inferred by MeerKAT PTA \cite{Miles2025MPTA}, providing an independent constraint on the candidate GWB signal. Our constraint also approaches the amplitude range suggested by recent radio PTA results (in 2023, gray region in Figure~\ref{fig:GWB_PTA_comparison}), demonstrating the increasing sensitivity of the GPTA. These results highlight the complementary role of gamma-ray pulsar timing in independently testing candidate nanohertz GWB signals reported by radio PTAs.

Previously, the sensitivity increase of the GPTA has been projected using the weak-signal scaling relation $A_{\rm GWB}\propto T_{\rm obs}^{-13/6}$ \cite{Siemens2013Scaling,FermiGPTA2022}, as illustrated by the dotted curve in Figure~\ref{fig:GWB_PTA_comparison}.
Here we instead fit the set of GPTA bounds with a power-law decline and find a function of $T_{\rm obs}^{-2.8}$, which is quicker than the weak-signal scaling relation (see the red solid curve in Figure~\ref{fig:GWB_PTA_comparison}). Consequently, 
our projected time for the GPTA to reach the common amplitude range inferred by NANOGrav/EPTA/PPTA/CPTA experiments in 2023 is mid-2027, about two years earlier than the projection based on DR1 in the case of weak-signal scaling.
This improvement highlights the gain in sensitivity achieved through more frequent sampling and the resulting larger number of TOAs. We emphasize that this estimate relies on an idealized weak-signal scaling relation and should therefore be regarded as a sensitivity projection rather than a prediction of a GWB detection.

Looking ahead, the GPTA sensitivity may be further improved by extending the observational baseline, optimizing the cadence, refining photon weighting and pulse-profile modeling, and incorporating additional gamma-ray millisecond pulsars. Continued observations will enable increasingly sensitive tests of the candidate GWB amplitude range using an independent gamma-ray data set. Finally, the proposed Very Large Area gamma-ray Space Telescope (VLAST), with its substantially larger effective area and acceptance than Fermi-LAT, offers considerable potential for high-precision gamma-ray pulsar timing and more sensitive searches for the nanohertz GWB \cite{VLAST2022,2024NuScT..35..149P}. Notably, the VLAST Pathfinder (VLAST-P), a technology-demonstration payload for the future mission, was successfully launched aboard the Pathfinder satellite on 11 November 2026 and is designed to demonstrate and verify key technologies in orbit. Our results significantly strengthen the GPTA constraints on the GWB, further demonstrating the potential of gamma-ray pulsar timing as an independent probe of the nanohertz gravitational-wave universe.

\textit{Acknowledgments.--} We thank the Fermi-LAT Collaboration for making the gamma-ray data used in this work publicly available and for their continued efforts in operating the Fermi Gamma-ray Space Telescope.
This work is supported by the Astrometric Reference Frame project (No. JZZX-020501), the National Natural Science Foundation of China  (Nos. 12588101, 12220101003), the Strategic Priority Research Program of the Chinese Academy of Sciences (No. XDB0550400), and the Project for Young Scientists in Basic Research of Chinese Academy of Sciences (No. YSBR-061). Y.Z.F thanks the support of New Cornerstone Science Foundation through the XPLORER PRIZE. 

\bibliographystyle{apsrev4-2}
\bibliography{reference}

\clearpage

\appendix
\section{Supplementary Material}

\setcounter{table}{0}
\renewcommand{\thetable}{S\arabic{table}}

\setcounter{equation}{0}
\renewcommand{\theequation}{S\arabic{equation}}

\subsection{A. TOA estimation}

We construct gamma-ray times of arrival (TOAs) by dividing the Fermi-LAT photon data into four datasets with fixed cadences of 122, 182, 243, and 365 days. Following the TOA estimation procedure used in Ref.~\cite{FermiGPTA2022},  we estimate the pulse phase within each time interval by fitting the photon phase distribution with a pulse-profile template constructed from the full observation span, and then derive one TOA for each valid interval.
The final time bin is retained even when its duration is shorter than the nominal cadence.
For each cadence, we then refit the timing model using the resulting TOAs and calculate the timing residuals. 

In this work, our initial sample consists of the 35 bright gamma-ray millisecond pulsars (MSPs) included in the first GPTA data release (DR1) \cite{FermiGPTA2022}. We use the corresponding timing models (par files) from GPTA DR1 and adopt the DE440 Solar System ephemeris \cite{Park2021DE440} for the timing analysis.

For each cadence, we independently exclude pulsars whose root-mean-square (RMS) of the post-fit timing residuals exceeds five times their median TOA uncertainty. This criterion is adopted to exclude pulsars with potentially unreliable TOAs or timing models that inadequately describe the full 17.6-year data span, as indicated by residual scatter substantially larger than expected from the TOA measurement uncertainties. We find that the number of pulsars retained decreases with decreasing cadence, and that the pulsar sample in each longer-cadence dataset contains all the pulsars retained in the corresponding shorter-cadence datasets. This trend may be attributed to the reduced number of photons per time bin at shorter cadences. For a given pulsar, statistical fluctuations in some time bins may become sufficiently large at shorter cadences to prevent reliable fitting of the pulse profile, resulting in unreliable TOA estimates. We therefore exclude such pulsars from the corresponding cadence dataset.

The median TOA uncertainties and post-fit timing-residual RMS values for the four cadence datasets are summarized in Tables~\ref{tab:noise_122d}, \ref{tab:noise_182d}, \ref{tab:noise_243d}, and \ref{tab:noise_365d}. With 17.6 years of Fermi-LAT observations, the most precise TOAs achieve uncertainties at the $\mu\mathrm{s}$ level, comparable to those obtained in the 12.5-year GPTA dataset \cite{FermiGPTA2022}.

\subsection{B. Single-pulsar noise analysis}
We perform Bayesian single-pulsar noise analyses using \texttt{enterprise} \cite{Enterprise2020}, with \texttt{PTMCMCSampler} \cite{Ellis2017PTMCMCSampler}. The noise model consists of a white-noise component and an optional intrinsic red-noise component. The white-noise model is parameterized by EFAC and EQUAD, with the effective TOA uncertainty given by
\begin{equation}
    \sigma_{\rm eff}^{2}
    = {\rm EFAC}^{2}\sigma_{\rm TOA}^{2}
    + {\rm EQUAD}^{2},
\end{equation}
where $\sigma_{\rm TOA}$ denotes the nominal TOA uncertainty. The intrinsic red noise is described by a power-law power spectral density (PSD),
\begin{equation}
    P_{\rm RN}(f)
    = \frac{A_{\rm RN}^{2}}{12\pi^{2}}
    \left(\frac{f}{f_{\rm yr}}\right)^{-\gamma_{\rm RN}}
    f_{\rm yr}^{-3},
\end{equation}
where $A_{\rm RN}$ and $\gamma_{\rm RN}$ are the dimensionless amplitude and spectral index of the intrinsic red noise, respectively, and $f_{\rm yr}=1\,{\rm yr}^{-1}$. We sample the logarithmic parameters $\log_{10} A_{\rm RN}$ and $\gamma_{\rm RN}$, together with EFAC and $\log_{10}{\rm EQUAD}$. The prior distributions adopted for the noise parameters are listed in Table~\ref{tab:priors}.

We use the Savage--Dickey (SD) density ratio to calculate Bayes factors for the inclusion of additional noise components \cite{Dickey1971,VerdinelliWasserman1995,Trotta2007SavageDickey}. Specifically, $\mathcal{B}_{\rm white}$ quantifies the evidence for introducing EFAC and EQUAD relative to the nominal TOA uncertainties, while $\mathcal{B}_{\rm red}$ quantifies the evidence for adding intrinsic red noise to the white-noise model. Noise parameters are reported only when the corresponding Bayes factor exceeds unity; otherwise, an em dash (---) is used. The reported parameter values correspond to the posterior medians and their associated $1\sigma$ uncertainties. The noise results for the four cadence datasets are summarized in Tables~\ref{tab:noise_122d}, \ref{tab:noise_182d}, \ref{tab:noise_243d}, and \ref{tab:noise_365d}.
Most pulsars do not show strong evidence for deviations from
the nominal TOA uncertainties in the white-noise model.
Specifically, $\mathcal{B}_{\rm white}>3$ for 7, 4, 3, and
2 pulsars in the 122-, 182-, 243-, and 365-day datasets,
respectively. The results for intrinsic red noise
are even less significant. Only J2034+3632 exhibits evidence
for an intrinsic red-noise component, with
$\mathcal{B}_{\rm red}>3$, and only in the 122-day dataset.

\subsection{C. Search for alternative common-spectrum stochastic processes}

In addition to the Hellings--Downs (HD)-correlated isotropic GWB model with a fixed spectral index of $\gamma=13/3$ for a population of inspiralling supermassive black hole (SMBH) binaries, we perform Bayesian searches for several alternative common-signal hypotheses.
These searches extend our investigation to GWB models with a broader range of spectral properties and alternative origins of common red signals:

\begin{itemize}
    \item An HD-correlated GWB with a freely varying spectral index
    $\gamma$, allowing for a broader range of possible astrophysical
    and cosmological GWB spectra. Such backgrounds may arise from sources beyond the SMBH binaries, including cosmic string networks, primordial tensor fluctuations generated during inflation, and other early-Universe processes such as first-order phase transitions and scalar-induced gravitational waves;
    
    \item An common uncorrelated red-noise process (CURN), in which all pulsars share the same red-noise spectrum but exhibit no spatial correlations. This model provides a phenomenological description of common-spectrum signals without assuming a specific spatial correlation pattern;

    \item A monopolar-correlated common process, which can arise
    from terrestrial clock errors or inaccuracies in time standards,
    producing common timing-residual variations across all pulsars;

    \item A dipolar-correlated common process, which is associated with potential Solar System ephemeris errors, particularly uncertainties in planetary positions and the resulting inaccuracies in the Solar System barycenter position.
\end{itemize}

For each common-signal model, we adopt a power-law power spectral density and perform separate analyses with and without BAYESEPHEM to account for potential Solar System ephemeris uncertainties. We compare the signal models using the SD Bayes factors. The corresponding Bayes factors and 95\% upper limits on the signal parameters are summarized in Table~\ref{tab:signal}.

Across all four cadence datasets, the Bayes factors remain below unity, ranging from 0.65 to 0.79, showing no preference for the tested common-signal models over the noise-only model. The results obtained with and without BAYESEPHEM are broadly consistent across the four cadence datasets.

The fixed-index HD-correlated GWB model ($\gamma=13/3$), corresponding to the standard background from a population of inspiralling SMBH binaries, yields Bayes factors ranging from 0.65 to 0.78, with relatively stable 95\% amplitude upper limits of $\log_{10} A^{95\%}=-14.43$ to $-14.24$.

The free-index HD model gives Bayes factors of 0.69--0.79 and 95\% amplitude upper limits of $\log_{10} A^{95\%}=-13.58$ to $-12.80$. Allowing the spectral index to vary provides a more general description of possible astrophysical and cosmological GWB sources, including cosmic strings and primordial gravitational waves from inflation.

The CURN model yields Bayes factors of 0.68--0.74 and 95\% amplitude upper limits of $\log_{10} A^{95\%}=-13.63$ to $-13.36$, with no significant preference for an uncorrelated common red-noise signal.

For the monopolar-correlated model, the Bayes factors range from 0.65 to 0.73, with 95\% amplitude upper limits of $\log_{10} A^{95\%}=-13.89$ to $-13.58$. These results show no preference for a monopolar common signal, which could be associated with terrestrial clock errors.

For the dipolar-correlated model, the Bayes factors range from
0.66 to 0.75, with 95\% amplitude upper limits of
$\log_{10} A^{95\%}=-13.73$ to $-13.24$. Given our adoption of the DE440 Solar System ephemeris, this search provides a complementary test for potential dipolar signals arising from ephemeris inaccuracies. No preference is found for such a signal.

\begin{table*}[htbp]
\caption{TOA estimation and single-pulsar noise analysis for the 122-day cadence dataset. The columns list the pulsar name, number of TOAs ($N_{\rm TOA}$), median TOA uncertainty ($\mu\mathrm{s}$), post-fit timing-residual RMS ($\mu\mathrm{s}$), and Bayes factors for the white- and red-noise model comparisons, $\mathcal{B}_{\rm white}$ and $\mathcal{B}_{\rm red}$, respectively. For cases with $\mathcal{B}>3$, the corresponding noise parameters are reported as posterior medians with $1\sigma$ uncertainties; em dashes indicate components with $\mathcal{B}\leq 3$.}
\label{tab:noise_122d}
\footnotesize
\setlength{\tabcolsep}{3.8pt}
\begin{tabular}{l|ccc|ccc|ccc}
\toprule
Pulsar & $N_{\rm TOA}$ & \shortstack{median TOA\\error ($\mu$s)} & \shortstack{post‑fit residual\\RMS ($\mu$s)} & $\mathcal{B}_{\rm white}$ & EFAC & $\log_{10}{\rm EQUAD}$ & $\mathcal{B}_{\rm red}$ & $\log_{10}A_{\rm RN}$ & $\gamma_{\rm RN}$ \\
\midrule
J0030+0451  & 53 & 5.14  & 4.91  & 0.06 & --- & --- & 0.77 & --- & --- \\
J0034-0534  & 53 & 17.74 & 30.60 & 2.73 & --- & --- & 0.83 & --- & --- \\
J0101-6422  & 53 & 11.14 & 16.93 & 8.49 & $1.41\pm0.16$ & $-7.19\pm1.23$ & 0.85 & --- & --- \\
J0102+4839  & 53 & 18.43 & 27.11 & 2.54 & --- & --- & 0.98 & --- & --- \\
J0340+4130  & 53 & 22.77 & 25.42 & 0.07 & --- & --- & 0.90 & --- & --- \\
J0418+6635  & 53 & 32.90 & 45.22 & 11.89 & $1.39\pm0.15$ & $-6.96\pm1.38$ & 1.06 & --- & --- \\
J0613-0200  & 53 & 24.09 & 21.19 & 0.10 & --- & --- & 0.84 & --- & --- \\
J0614-3329  & 53 & 2.92  & 3.39  & 0.26 & --- & --- & 1.11 & --- & --- \\
J1124-3653  & 53 & 12.62 & 26.56 & $>100$ & $2.17\pm0.25$ & $-6.81\pm1.45$ & 1.04 & --- & --- \\
J1231-1411  & 53 & 3.19  & 3.23  & 0.06 & --- & --- & 0.75 & --- & --- \\
J1514-4946  & 53 & 14.23 & 12.55 & 0.10 & --- & --- & 0.85 & --- & --- \\
J1536-4948  & 53 & 9.35  & 7.86  & 0.09 & --- & --- & 0.89 & --- & --- \\
J1614-2230  & 53 & 8.33  & 15.82 & 78.42 & $1.49\pm0.15$ & $-7.30\pm1.17$ & 0.87 & --- & --- \\
J1625-0021  & 53 & 24.38 & 27.52 & 0.12 & --- & --- & 0.92 & --- & --- \\
J1858-2216  & 53 & 23.51 & 71.98 & $>100$ & $2.71\pm0.29$ & $-7.09\pm1.30$ & 0.96 & --- & --- \\
J1902-5105  & 53 & 10.47 & 12.15 & 0.08 & --- & --- & 0.81 & --- & --- \\
J2017+0603  & 53 & 17.76 & 19.83 & 0.07 & --- & --- & 0.89 & --- & --- \\
J2034+3632  & 53 & 44.50 & 129.36 & $>100$ & $3.76\pm0.40$ & $-6.91\pm1.41$ & $>100$ & $-10.49\pm0.05$ & $0.31\pm0.25$ \\
J2043+1711  & 53 & 9.89  & 12.09 & 0.07 & --- & --- & 1.02 & --- & --- \\
J2241-5236  & 53 & 7.76  & 18.86 & $>100$ & $2.26\pm0.25$ & $-7.20\pm1.23$ & 0.82 & --- & --- \\
J2302+4442  & 53 & 11.79 & 12.35 & 0.07 & --- & --- & 0.83 & --- & --- \\
\bottomrule
\end{tabular}
\end{table*}

\begin{table*}[htbp]
\caption{Same as Table~\ref{tab:noise_122d}, for the 182‑day cadence dataset.}
\label{tab:noise_182d}
\footnotesize
\setlength{\tabcolsep}{3.8pt}
\begin{tabular}{l|ccc|ccc|ccc}
\toprule
Pulsar & $N_{\rm TOA}$ & \shortstack{median TOA\\error ($\mu$s)} & \shortstack{post‑fit residual\\RMS ($\mu$s)} & $\mathcal{B}_{\rm white}$ & EFAC & $\log_{10}{\rm EQUAD}$ & $\mathcal{B}_{\rm red}$ & $\log_{10}A_{\rm RN}$ & $\gamma_{\rm RN}$ \\
\midrule
J0030+0451 & 36 & 4.14 & 4.72 & 0.11 & --- & --- & 0.77 & --- & --- \\
J0034-0534 & 36 &14.19 &14.16 & 0.07 & --- & --- & 0.90 & --- & --- \\
J0101-6422 & 36 & 9.67 &16.27 & $>100$ & $1.80\pm0.27$ & $-7.17\pm1.24$ & 0.86 & --- & --- \\
J0102+4839 & 36 &16.01 &15.51 & 0.07 & --- & --- & 0.95 & --- & --- \\
J0312-0921 & 36 &13.70 &57.44 & $>100$ & $4.60\pm0.36$ & $-6.38\pm1.50$ & 1.34 & --- & --- \\
J0340+4130 & 36 &18.89 &19.44 & 0.07 & --- & --- & 0.91 & --- & --- \\
J0418+6635 & 36 &26.75 &36.25 & 0.46 & --- & --- & 0.98 & --- & --- \\
J0613-0200 & 36 &19.74 &11.86 & 19.45 & $0.62\pm0.08$ & $-7.01\pm1.35$ & 0.82 & --- & --- \\
J0614-3329 & 36 & 2.42 & 2.70 & 0.16 & --- & --- & 1.68 & --- & --- \\
J0740+6620 & 36 &11.68 &11.89 & 0.07 & --- & --- & 0.97 & --- & --- \\
J1124-3653 & 36 &10.96 &10.50 & 0.08 & --- & --- & 0.84 & --- & --- \\
J1231-1411 & 36 & 2.55 & 2.47 & 0.09 & --- & --- & 0.74 & --- & --- \\
J1514-4946 & 36 &11.66 & 9.80 & 0.08 & --- & --- & 0.89 & --- & --- \\
J1536-4948 & 36 & 7.60 & 6.13 & 0.08 & --- & --- & 0.86 & --- & --- \\
J1614-2230 & 36 & 7.15 & 6.78 & 0.07 & --- & --- & 0.79 & --- & --- \\
J1625-0021 & 36 &20.34 &19.46 & 0.07 & --- & --- & 0.99 & --- & --- \\
J1630+3734 & 36 & 5.92 & 7.54 & 0.11 & --- & --- & 0.82 & --- & --- \\
J1858-2216 & 36 &17.64 &24.12 & 0.10 & --- & --- & 1.13 & --- & --- \\
J1902-5105 & 36 & 8.59 & 6.95 & 0.09 & --- & --- & 0.92 & --- & --- \\
J1939+2134 & 36 & 8.33 &10.99 & 0.13 & --- & --- & 0.84 & --- & --- \\
J2017+0603 & 36 &14.62 &16.77 & 0.14 & --- & --- & 0.88 & --- & --- \\
J2034+3632 & 36 &38.07 &34.76 & 0.07 & --- & --- & 0.92 & --- & --- \\
J2043+1711 & 36 & 8.41 &11.97 & 0.17 & --- & --- & 0.90 & --- & --- \\
J2241-5236 & 36 & 6.65 &18.22 & $>100$ & $2.63\pm0.37$ & $-7.27\pm1.17$ & 1.15 & --- & --- \\
J2302+4442 & 36 & 9.78 &10.23 & 0.07 & --- & --- & 0.81 & --- & --- \\
\bottomrule
\end{tabular}
\end{table*}

\begin{table*}[htbp]
\caption{Same as Table~\ref{tab:noise_122d}, for the 243‑day cadence dataset.}
\label{tab:noise_243d}
\footnotesize
\setlength{\tabcolsep}{3.8pt}
\begin{tabular}{l|ccc|ccc|ccc}
\toprule
Pulsar & $N_{\rm TOA}$ & \shortstack{median TOA\\error ($\mu$s)} & \shortstack{post‑fit residual\\RMS ($\mu$s)} & $\mathcal{B}_{\rm white}$ & EFAC & $\log_{10}{\rm EQUAD}$ & $\mathcal{B}_{\rm red}$ & $\log_{10}A_{\rm RN}$ & $\gamma_{\rm RN}$ \\
\midrule
J0030+0451 & 27 & 3.56 & 3.34 & 0.09 & --- & --- & 0.77 & --- & --- \\
J0034-0534 & 27 &12.22 &11.09 & 0.09 & --- & --- & 0.84 & --- & --- \\
J0101-6422 & 27 & 8.55 & 6.65 & 0.10 & --- & --- & 0.97 & --- & --- \\
J0102+4839 & 27 &13.32 &14.28 & 0.11 & --- & --- & 0.95 & --- & --- \\
J0312-0921 & 27 &12.89 &30.87 & $>100$ & $3.02\pm0.56$ & $-7.00\pm1.36$ & 0.96 & --- & --- \\
J0340+4130 & 27 &16.99 &17.76 & 0.08 & --- & --- & 0.90 & --- & --- \\
J0418+6635 & 27 &22.63 &31.12 & 0.11 & --- & --- & 0.91 & --- & --- \\
J0613-0200 & 27 &16.96 &14.90 & 0.11 & --- & --- & 0.83 & --- & --- \\
J0614-3329 & 27 & 2.08 & 2.66 & 0.44 & --- & --- & 1.13 & --- & --- \\
J0740+6620 & 27 & 9.78 & 9.36 & 0.10 & --- & --- & 0.95 & --- & --- \\
J1124-3653 & 27 &10.03 &13.74 & 0.30 & --- & --- & 0.91 & --- & --- \\
J1231-1411 & 27 & 2.20 & 2.58 & 0.09 & --- & --- & 0.72 & --- & --- \\
J1513-2550 & 27 &15.61 &21.75 & 1.53 & --- & --- & 1.00 & --- & --- \\
J1514-4946 & 27 &10.38 & 7.07 & 0.11 & --- & --- & 0.96 & --- & --- \\
J1536-4948 & 27 & 6.48 & 4.00 & 0.15 & --- & --- & 0.92 & --- & --- \\
J1614-2230 & 27 & 6.50 & 7.06 & 0.10 & --- & --- & 0.78 & --- & --- \\
J1625-0021 & 27 &17.98 &18.82 & 0.13 & --- & --- & 0.92 & --- & --- \\
J1630+3734 & 27 & 5.35 & 7.32 & 0.61 & --- & --- & 0.82 & --- & --- \\
J1741+1351 & 27 &16.66 &35.04 & $>100$ & $1.87\pm0.28$ & $-7.06\pm1.31$ & 0.91 & --- & --- \\
J1816+4510 & 27 &15.13 &20.11 & 0.08 & --- & --- & 0.88 & --- & --- \\
J1858-2216 & 27 &15.40 &14.97 & 0.11 & --- & --- & 0.85 & --- & --- \\
J1902-5105 & 27 & 7.51 & 5.86 & 0.11 & --- & --- & 0.86 & --- & --- \\
J1939+2134 & 27 & 7.51 & 9.60 & 0.19 & --- & --- & 0.87 & --- & --- \\
J2017+0603 & 27 &12.85 &12.60 & 0.08 & --- & --- & 0.90 & --- & --- \\
J2034+3632 & 27 &31.77 &24.34 & 0.15 & --- & --- & 0.99 & --- & --- \\
J2043+1711 & 27 & 7.40 &13.12 & 0.12 & --- & --- & 0.98 & --- & --- \\
J2214+3000 & 27 &14.96 &18.62 & 0.08 & --- & --- & 0.88 & --- & --- \\
J2241-5236 & 27 & 5.55 &15.99 & $>100$ & $2.65\pm0.47$ & $-7.22\pm1.20$ & 0.85 & --- & --- \\
J2302+4442 & 27 & 8.65 & 8.80 & 0.08 & --- & --- & 0.83 & --- & --- \\
\bottomrule
\end{tabular}
\end{table*}

\begin{table*}[htbp]
\caption{Same as Table~\ref{tab:noise_122d}, for the 365‑day cadence dataset.}
\label{tab:noise_365d}
\footnotesize
\setlength{\tabcolsep}{3.8pt}
\begin{tabular}{l|ccc|ccc|ccc}
\toprule
Pulsar & $N_{\rm TOA}$ & \shortstack{median TOA\\error ($\mu$s)} & \shortstack{post‑fit residual\\RMS ($\mu$s)} & $\mathcal{B}_{\rm white}$ & EFAC & $\log_{10}{\rm EQUAD}$ & $\mathcal{B}_{\rm red}$ & $\log_{10}A_{\rm RN}$ & $\gamma_{\rm RN}$ \\
\midrule
J0030+0451 & 18 & 2.93 & 2.92 & 0.12 & --- & --- & 0.77 & --- & --- \\
J0034-0534 & 18 & 9.79 & 7.31 & 0.15 & --- & --- & 0.84 & --- & --- \\
J0101-6422 & 18 & 7.38 & 3.48 & 0.19 & --- & --- & 1.07 & --- & --- \\
J0102+4839 & 18 &11.56 &11.64 & 0.13 & --- & --- & 0.88 & --- & --- \\
J0312-0921 & 18 &11.18 &34.72 & $>100$ & $4.23\pm0.59$ & $-6.98\pm1.36$ & 1.15 & --- & --- \\
J0340+4130 & 18 &13.69 &11.49 & 0.11 & --- & --- & 0.95 & --- & --- \\
J0418+6635 & 18 &19.66 &15.84 & 0.12 & --- & --- & 1.04 & --- & --- \\
J0533+6759 & 18 &14.28 & 8.55 & 0.28 & --- & --- & 0.93 & --- & --- \\
J0613-0200 & 18 &14.07 & 7.26 & 2.25 & --- & --- & 0.83 & --- & --- \\
J0614-3329 & 18 & 1.72 & 2.26 & 0.69 & --- & --- & 0.99 & --- & --- \\
J0740+6620 & 18 & 7.95 & 8.32 & 0.11 & --- & --- & 0.92 & --- & --- \\
J1124-3653 & 18 & 7.62 & 6.17 & 0.13 & --- & --- & 0.91 & --- & --- \\
J1231-1411 & 18 & 1.82 & 1.60 & 0.11 & --- & --- & 0.74 & --- & --- \\
J1513-2550 & 18 &13.19 &13.32 & 0.17 & --- & --- & 1.61 & --- & --- \\
J1514-4946 & 18 & 8.24 & 6.35 & 0.27 & --- & --- & 1.07 & --- & --- \\
J1536-4948 & 18 & 5.27 & 3.39 & 0.22 & --- & --- & 1.03 & --- & --- \\
J1543-5149 & 18 &18.71 &19.98 & 0.18 & --- & --- & 0.95 & --- & --- \\
J1614-2230 & 18 & 4.91 & 4.49 & 0.11 & --- & --- & 0.80 & --- & --- \\
J1625-0021 & 18 &15.45 &15.26 & 0.13 & --- & --- & 0.91 & --- & --- \\
J1630+3734 & 18 & 4.27 & 5.40 & 0.21 & --- & --- & 0.87 & --- & --- \\
J1741+1351 & 18 &14.81 &23.03 & 0.89 & --- & --- & 0.86 & --- & --- \\
J1810+1744 & 18 & 8.25 & $<0.01$ & 1.00 & --- & --- & 1.40 & --- & --- \\
J1816+4510 & 18 &12.70 &15.81 & 0.14 & --- & --- & 0.90 & --- & --- \\
J1858-2216 & 18 &12.49 &10.61 & 0.13 & --- & --- & 0.94 & --- & --- \\
J1902-5105 & 18 & 6.16 & 4.08 & 0.19 & --- & --- & 1.01 & --- & --- \\
J1939+2134 & 18 & 6.14 & 6.88 & 0.15 & --- & --- & 0.89 & --- & --- \\
J2017+0603 & 18 &10.42 & 9.89 & 0.10 & --- & --- & 0.88 & --- & --- \\
J2034+3632 & 18 &26.79 &19.75 & 0.12 & --- & --- & 0.97 & --- & --- \\
J2043+1711 & 18 & 5.94 & 8.09 & 0.18 & --- & --- & 0.91 & --- & --- \\
J2214+3000 & 18 &13.56 &16.17 & 0.30 & --- & --- & 0.90 & --- & --- \\
J2241-5236 & 18 & 4.67 &12.41 & $>100$ & $2.90\pm0.70$ & $-7.20\pm1.23$ & 2.90 & --- & --- \\
J2302+4442 & 18 & 6.95 & 7.26 & 0.12 & --- & --- & 0.85 & --- & --- \\
\bottomrule
\end{tabular}
\end{table*}

\begin{table*}[htbp]
\caption{Prior distributions used in the Bayesian noise and common‑signal analyses.
All free parameters have uniform priors over the specified intervals.
The SMBH-binary GWB spectral index is fixed at $\gamma=13/3$.
Noise parameters are specified on a per-pulsar basis, whereas common-signal parameters are shared by all pulsars within each PTA.}
\label{tab:priors}
\footnotesize
\setlength{\tabcolsep}{6pt}
\begin{tabular}{lll}
\toprule
Parameter & Symbol & Uniform prior range \\
\midrule
\multicolumn{3}{c}{\bf{Noise}} \\
\midrule
TOA error scaling & $\rm EFAC$ & $[0.1,\,5.0]$ \\
Additive white‑noise term & $\log_{10}{\rm EQUAD}$ & $[-9.0,\,-5.0]$ \\
Intrinsic red‑noise amplitude & $\log_{10}A_{\rm RN}$ & $[-20.0,\,-10.0]$ \\
Intrinsic red‑noise spectral index & $\gamma_{\rm RN}$ & $[0.0,\,7.0]$ \\
\midrule
\multicolumn{3}{c}{\bf{Common signal}} \\
\midrule
Common‑process amplitude & $\log_{10}A$ & $[-20.0,\,-10.0]$ \\
Common‑process spectral index (free) & $\gamma$ & $[0.0,\,7.0]$ \\
Fixed SMBH‑binary GWB spectral index & $\gamma$ & $4.33$ (fixed, $\gamma=13/3$) \\
\bottomrule
\end{tabular}
\end{table*}

\begin{table*}[t]
\caption{Bayes factors and 95\% upper limits for the common‑signal
hypotheses, obtained with and without \textsc{BayesEphem}. The Bayes factor
$\mathcal{B}$ refers to the signal model against the noise-only model; upper limits are one‑sided 95th posterior percentiles.
The physically motivated Hellings-Downs (HD) model adopts a free spectral index.
For the GWB row the spectral index is fixed to $\gamma=13/3$, hence no upper‑limit is quoted for $\gamma$.
Note that $\gamma$ upper limits for free‑index models are prior‑dominated.}
\label{tab:signal}
\footnotesize
\setlength{\tabcolsep}{4pt}
\begin{tabular}{llcccccc}
\toprule
& & \multicolumn{3}{c}{With BAYESEPHEM} & \multicolumn{3}{c}{Without BAYESEPHEM} \\
\cmidrule(lr){3-5}\cmidrule(lr){6-8}
Cadence & Model & $\mathcal{B}$ & $\log_{10}A^{95\%}$ & $\gamma^{95\%}$ & $\mathcal{B}$ & $\log_{10}A^{95\%}$ & $\gamma^{95\%}$ \\
\midrule
\multirow{5}{*}{122\,d} & \textbf{GWB (fixed $\gamma=13/3$)} & \textbf{0.65} & $\mathbf{-14.43}$ & \textbf{---} & \textbf{0.66} & $\mathbf{-14.41}$ & \textbf{---} \\
 & HD (free $\gamma$) & 0.69 & $-13.58$ & 6.54 & 0.71 & $-13.57$ & 6.55 \\
 & CURN & 0.70 & $-13.63$ & 6.54 & 0.71 & $-13.63$ & 6.61 \\
 & Monopole & 0.70 & $-13.77$ & 6.52 & 0.73 & $-13.72$ & 6.53 \\
 & Dipole & 0.67 & $-13.73$ & 6.53 & 0.67 & $-13.69$ & 6.47 \\
\midrule
\multirow{5}{*}{182\,d} & \textbf{GWB (fixed $\gamma=13/3$)} & \textbf{0.69} & $\mathbf{-14.30}$ & \textbf{---} & \textbf{0.78} & $\mathbf{-14.25}$ & \textbf{---} \\
 & HD (free $\gamma$) & 0.72 & $-13.48$ & 6.57 & 0.79 & $-13.39$ & 6.59 \\
 & CURN & 0.71 & $-13.52$ & 6.53 & 0.74 & $-13.45$ & 6.54 \\
 & Monopole & 0.65 & $-13.89$ & 6.50 & 0.65 & $-13.87$ & 6.58 \\
 & Dipole & 0.72 & $-13.27$ & 6.51 & 0.75 & $-13.24$ & 6.54 \\
\midrule
\multirow{5}{*}{243\,d} & \textbf{GWB (fixed $\gamma=13/3$)} & \textbf{0.73} & $\mathbf{-14.24}$ & \textbf{---} & \textbf{0.74} & $\mathbf{-14.25}$ & \textbf{---} \\
 & HD (free $\gamma$) & 0.75 & $-13.47$ & 6.54 & 0.75 & $-13.45$ & 6.57 \\
 & CURN & 0.69 & $-13.60$ & 6.55 & 0.71 & $-13.54$ & 6.49 \\
 & Monopole & 0.67 & $-13.78$ & 6.56 & 0.68 & $-13.79$ & 6.53 \\
 & Dipole & 0.69 & $-13.57$ & 6.48 & 0.71 & $-13.54$ & 6.51 \\
\midrule
\multirow{5}{*}{365\,d} & \textbf{GWB (fixed $\gamma=13/3$)} & \textbf{0.66} & $\mathbf{-14.36}$ & \textbf{---} & \textbf{0.68} & $\mathbf{-14.31}$ & \textbf{---} \\
 & HD (free $\gamma$) & 0.76 & $-12.80$ & 6.51 & 0.75 & $-13.21$ & 6.51 \\
 & CURN & 0.68 & $-13.46$ & 6.60 & 0.71 & $-13.36$ & 6.50 \\
 & Monopole & 0.70 & $-13.58$ & 6.49 & 0.69 & $-13.63$ & 6.40 \\
 & Dipole & 0.66 & $-13.64$ & 6.46 & 0.70 & $-13.49$ & 6.54 \\
\bottomrule
\end{tabular}
\end{table*}

\end{document}